# Passively mode-locked diode-pumped Nd:YVO$_4$ oscillator operating at ultra-low repetition rate

D. N. Papadopoulos, S. Forget, M. Delaigue, F. Druon, F. Balembois and P. Georges

Laboratoire Charles Fabry de l'Institut d'Optique, UMR 8501 du CNRS
Université Paris-Sud, 91403 Orsay, France

We demonstrate the operation of an ultra low repetition rate, high peak power, picosecond diode pumped Nd:YVO$_4$ passively mode locked laser oscillator. Repetition rates even below 1 MHz were achieved with the use of a new design multiple-pass cavity and a semiconductor saturable absorber. Long term stable operation at 1.2 MHz, pulse duration of 16.3 ps and average output power of 470 mW corresponding to 24 KW peak power pulses is reported. This is, to our knowledge, the lowest repetition rate high peak power pulses ever generated directly from a picosecond laser resonator without cavity dumping.

Many applications in biology such as fluorescence lifetime measurements require pulsed laser sources emitting visible radiations. As the fluorescence lifetime of numerous molecules is in the hundreds of picoseconds range, short pulses around some tenths of picoseconds are ideally needed from reliable laser sources. The repetition rate of the laser source used is also a very important parameter in experiments using photon counting detection chains: high-repetition rates ensure a fast acquisition process of the fluorescence decay signal and allow to study of dynamic processes. However, if the repetition rate exceeds several MHZ, some problems appear in the complex signal processing devices during the acquisition of the data. Consequently a repetition rate around 1 or 2 MHz seems to be a good requirement for this kind of applications. A classical approach to generate short visible pulses is to use an oscillator operating at a fixed wavelength in the IR (1064 nm) and to use non-linear processes to reach the visible range. But producing picosecond pulses in the IR at a repetition rate of 1 MHz is not straightforward: a first alternative is the development of a Q-switched microchip laser.[1] However, in such systems pulse durations below 100 ps with high enough peak power cannot be directly achieved without the use of amplification stages. The second alternative is the development of a mode-locked laser oscillator. However, those sources operate typically at several tenths of MHz if the cavity size doesn't exceed some meters : for a mode-locked laser the repetition rate, f, of the pulses is fixed by the length of the cavity, L, according to the relation f=c/2L, where c is the speed of light. Therefore to achieve a repetition rate down to 1 MHz, the cavity length should be increased up to 150 meters. To achieve such long cavities one has to introduce a multiple pass cavity (MPC). In this letter we report the operation of a passively mode locked diode pumped Nd:YVO$_4$ oscillator at repetition rate even below 1MHz, emitting pulses of 16.3 ps duration with 470 mW average power.

The classical two concave mirrors Herriott-style MPC, already widely used in a number of applications, especially as a delay line as well as for gas cells,[2] has been recently used successfully for the decrease of the frequency of a Ti:Al$_2$O$_3$ mode locked laser down to 4 MHz.[3] In such a MPC however, the number of passes through the cavity is strictly defined by the distance between the two concave mirrors, while the beam entrance conditions i.e. the initial distance r of the beam from the axis of the MPC and the angle θ, only affects the shape and the size of the spot pattern on the concave mirrors.[2]

The configuration of the MPC used in our system is in fact the folded in two, with the help of two plane mirrors (Fig. 1), version of the Herriott-style MPC. In our configuration the distance of both concave mirrors (R=2 m) from the two plane mirrors is fixed at approximately the focal length of the concave mirrors, specifically at 105 cm. The beam is periodically focused and defocused after each reflection on the concave and the plane mirrors respectively. The over all ABCD matrix of the MPC in this case is of course no longer a unity transformation matrix.[3] Thus, the design of the cavity as a whole, already containing the MPC, is necessary. With the use of an ABCD matrix simulation program (simulating the MPC as a repeated series of 2 m concave mirrors separated by a distance of about 2 m) we saw that no significant change in the intracavity beam properties is observed when the number of passes through the MPC is arbitrarily increased, if the design starts with a small cavity (without the MPC) providing a fairly collimated beam. The main advantage of this MPC configuration is the almost continually controlled number of the passes through the MPC by simply changing the alignment of the plane and/or the concave mirrors. The remarkable freedom of the easy control on the beam path in the MPC, is the result of the use of the two plane mirrors. Each reflection on them, when the proper alignment is achieved, actually resets the beam into the MPC in such a way that forces it to stay in the MPC for an arbitrary large number of round trips. Possibly the use of three or even four instead of two plane mirrors could supply better and easier control on the number of reflections, complicating of course the final setup. Additionally, in our case it was of course not necessary to drill or cut the concave mirrors for the entrance and the exit of the beam.[3]

In this system a 5mm long, 0.1% doped Nd:YVO$_4$ crystal is pumped by a fiber coupled laser diode at 808 nm of 15 W maximum power. Passive mode locking is achieved with the use of a semiconductor saturable absorber mirror (SESAM)[4,5] with 6.1% modulation depth. During the alignment of the MPC inside the laser cavity, various repeated beam spot numbers and patterns were observed resulting each time the operation of the system on a number of different frequencies. Specifically, stable operation in 20, 12, 5, 3, 2.3 and finally 1.2 MHz was achieved by simply adjusting the alignment of the MPC. For the case of 1.2 MHz an asymmetric repeated elliptical beam spot pattern on the two concave mirrors (Fig. 2) and the corresponding on the two planes provided a cavity length of around 121 m after 56 round trips inside the MPC. Stable, single mode operation at repetition rate as low as 1.2 MHz, pulse duration of 16.3 ps duration with average power of 470 mW corresponding to 392 nJ pulse energy and 24 kW peak power is reported. The beam quality is close to the diffraction limit, with M$^2$ slightly above 1.1. In figure 3 is shown the photograph of the pulse train as measured with the help of a fast photodiode. Small peaks appearing just after the pulses correspond to relaxation peaks of the photodiode. In figure 4 are shown the autocorrelation trace of the pulses and the respective Gaussian fit curve. The FWHM of the Gaussian curve is 23 ps corresponding 16.3 ps pulse duration. The stability of the system for continuous operation during some weeks period was very satisfactory.

The output power was seriously confined by diffraction losses at the edges of the MPC mirrors. Replacing the SESAM with a second output mirror of the same transmission as the first one and by comparison of the power of the two outputs, the overall losses in the MPC (including the diffraction losses) were measured to be 30%. At higher frequencies i.e. when fewer passes through the MPC and simpler beam spot patterns on the mirrors were chosen, higher output powers of several Watts were possible (2.5 W for f = 2,3 MHz). On the other side, operation even below 1 MHz was also observed at 884 and 650 kHz. However in those frequencies severe multiple pulsing instabilities did not allow a long term operation. Nevertheless, we strongly believe that the reduction of the frequencies far below the 1 MHz is quite manageable with the use of this MPC configuration, if the design of the cavity is done from the beginning for such a purpose.

In such low frequencies and high intracavity energies and peak powers, careful design is crucial to avoid malfunction of the SESAM. The value of the modulation depth of the SESAM $\Delta R$, the tendency for Q-switching mode locking (QML) and multiple pulsing instabilities that the SESAM introduces and of course its damage threshold are the key points towards the correct design of the cavity. Generally in passively mode locked solid state lasers with weak gain saturation, such as an $Nd:YVO_4$ laser, an initial fluctuation of the intracavity power into the cw regime will monotonically increase until steady state mode locking is reached if and only if:[6]

$$\kappa P \ln(m_i) > T_r/T_c \qquad (1)$$

where $m_i$ represents the number of the initially oscillating modes, P could be approximated by the steady state intracavity power, $T_r$ is the round-trip time, $T_c$ the effective correlation time defined by the inverse 3 dB full width $\Delta \nu_{3dB}$ of the first beat note of the free-running laser and $\kappa$ a characteristic of the nonlinear device (SESAM), that gives the change in round-trip power gain per unit intracavity power.[6] As the values of P, $m_i$ and $T_c$ should be considered as fixed for a specific system, for very long cavities i.e. large $T_r$ values, the previous condition becomes very demanding. Since the left hand factor of (1) approximates the effective modulation depth, $\Delta R$,[6] of the SESAM, a choice of a sufficiently large value for $\Delta R$ is crucial. For $T_r \approx 1$ μs and typical values of the $\Delta \nu_{3dB}$ around 1-10 KHz[6] a secure choice of $\Delta R$ is above 3%. Our systems employs a SESAM with $\Delta R = 6.1\%$. On the other hand a SESAM introduces a Q-switching tendency that can drive the laser into the unwanted regime of Q-switching mode locking (QML). To obtain free of QML instabilities cw mode locking the intracavity pulse energy $E_p$ should satisfy the condition:[7]

$$E_p > (F_{sat,L} A_{eff,L} F_{sat,A} A_{eff,A} \Delta R)^{1/2} \qquad (2)$$

where $F_{sat,L}$, $F_{sat,A}$ are the saturation fluence of the gain medium and of the SESAM respectively and $A_{eff,L}$, $A_{eff,A}$ the effective laser mode in the gain medium and on the SESAM respectively. Thus, for a choice of large $\Delta R$ and for the fixed values of $F_{sat,L}$ and $F_{sat,A}$ the design of the cavity should provide small enough values for the $A_{eff,L}$ and $A_{eff,A}$ and enough $E_p$, so that the condition (2) is satisfied too. However, the saturation of the SESAM i.e. the value of the factor $S = E_p/F_{sat,A} A_{eff,A}$, should be kept always below 20 to avoid multiple pulsing instabilities or even damage of the SESAM[8,9]. By proper choice of the pumping focusing optics and of the cavity mirrors we were able in some extent to achieve the proper values for $A_{eff,L}$ and $A_{eff,A}$ so that all previous conditions were simultaneously met. Thus, for the case of 1.2 MHz system i.e. for P=470 mW with a 10% OC corresponding to $E_p$=3.5 μJ, for a double pass (under strong overlapping of the beams) through the $Nd:YVO_4$ crystal i.e. $F_{sat,L} = h\nu/4\sigma_L = 30$ mJ/cm² and for a SESAM with $\Delta R = 6.1\%$ and $F_{sat,A} = 70$ μJ/cm² we have chosen an 1:1 two lens focusing of the pump beam, providing a mode area in the crystal estimated around $A_{eff,L} = 1.3 \cdot 10^{-3}$ cm² (pump fiber core diameter = 400 μm) and a 2 m focusing mirror on the SESAM providing an 770 μm mode diameter on the SESAM corresponding to $A_{eff,A}$=4.7 $10^{-3}$ cm$^2$. For the above values condition (2) is well satisfied since 3.5>0.88 and the saturation of the SESAM is S=10.

In conclusion, operation of an ultra low repetition rate, even below 1 MHz, mode locked Nd:YVO$_4$ oscillator has been demonstrated with a simple and low cost design. Stable long term operation in 1.2 MHz, pulse duration of 16.3 ps, and 24 kW peak power was achieved. For the best of our knowledge this is the lowest repetition rate ever reported for a completely passively mode locked laser with no use of complicated and expensive techniques such as cavity dumping. In preliminary experiments on frequency doubling, conversion efficiency into 532 nm as high as 50% was achieved with the use of 10 mm long periodically poled KTP crystal. These results are very promising for the further use of this source for the generation of short visible pulses for fluorescence lifetime measurements.

This research was partially supported by the research program POLA from the Contrat Plan Etat Région (2000-2006) (French State and Conseil Général de l'Essonne). D. N. Papadopoulos worked for this project under a Marie Curie fellowship. His current affiliation is: *Lasers and Applications Group, Physics Department, National Technical University of Athens, Heroon Polytechniou 9, 15780 Athens, Greece.*
The e-mail address of Sébastien Forget is: *sebastien.forget@iota.u-psud.fr*.

**References including the titles**

**Figures**

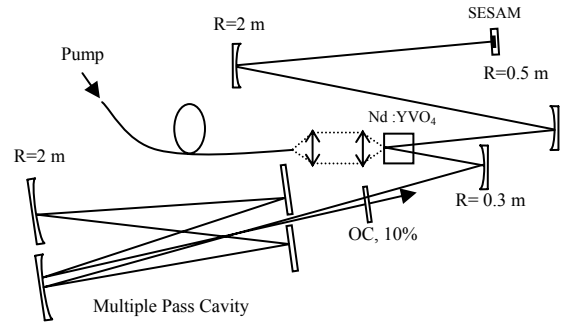

Fig. 1. Schematic of the laser cavity with the four low losses R>99.9 mirrors (two folding plane mirrors and two concave with R=2m) multiple-pass cavity.

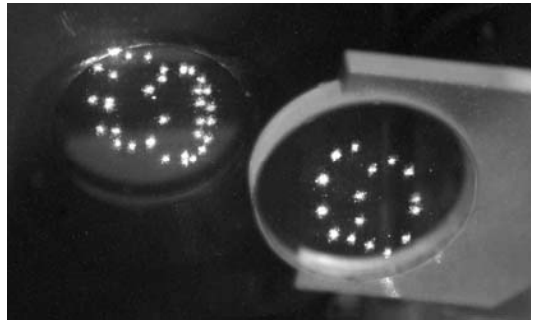

Fig. 2. Photograph of the MPC concave mirrors with the beam spot pattern on them for the 1.2 MHz repetition rate mode locked Nd:YVO$_4$ laser.

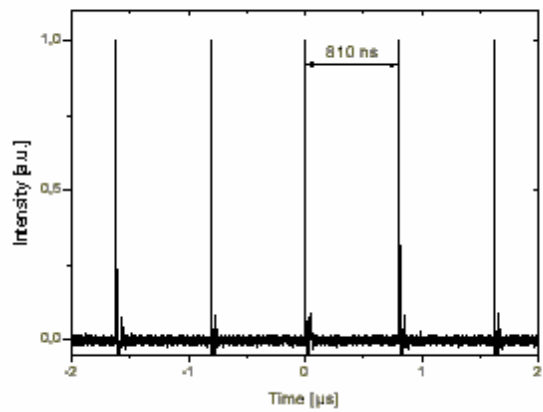

Fig. 3. Oscilloscope trace of a fast photodiode, showing the mode locked laser pulse train of 1.2 MHz of 810 ns separation.

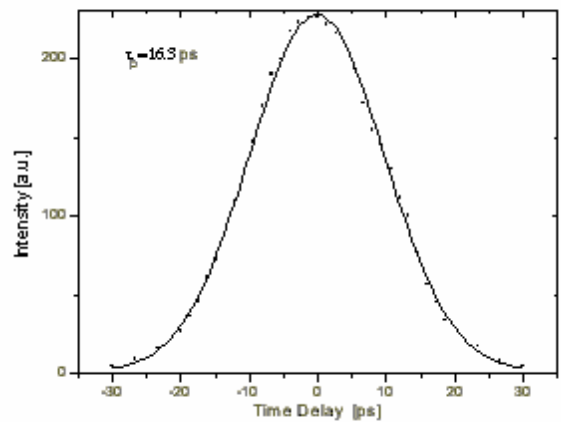

Fig. 4. Intensity autocorrelation trace showing a pulse duration of 16.3 ps assuming a Gaussian intensity profile.

**Figures captions**

Fig. 1. Schematic of the laser cavity with the four low losses R>99.9 mirrors (two folding plane mirrors and two concave with R=2m) multiple-pass cavity.

Fig. 2. Photograph of the MPC concave mirrors with the beam spot pattern on them for the 1.2 MHz repetition rate mode locked Nd:YVO$_4$ laser.

Fig. 3. Oscilloscope trace of a fast photodiode, showing the mode locked laser pulse train of 1.2 MHz of 810 ns separation.

Fig. 4. Intensity autocorrelation trace showing a pulse duration of 16.3 ps assuming a Gaussian intensity profile.